\documentclass[reprint, aps,prb,amsmath, amssymb, amsfonts,superscriptaddress]{revtex4-2}
\usepackage{graphicx}
\usepackage{xcolor}
\usepackage{cancel}
\usepackage{braket}
\usepackage{calc}
\usepackage{bbold}
\usepackage{bm}
\usepackage{soul}

\usepackage[colorlinks=true,citecolor=magenta,linkcolor=red,linktocpage=true,pagebackref=false]{hyperref}

\newtheorem{problem}{Problem}

\begin{document}

\title{
Characterizing and Mitigating Timing Noise-Induced Decoherence 
\\in Single Electron Sources
}

\author{Sungguen Ryu}
\email{sungguen@ifisc.uib-csic.es}
\affiliation{Institute for Cross-Disciplinary Physics and Complex Systems IFISC (UIB-CSIC), E-07122 Palma de Mallorca, Spain}
\author{Rosa L\'{o}pez}
\affiliation{Institute for Cross-Disciplinary Physics and Complex Systems IFISC (UIB-CSIC), E-07122 Palma de Mallorca, Spain}
\author{Lloren\c{c} Serra}
\affiliation{Institute for Cross-Disciplinary Physics and Complex Systems IFISC (UIB-CSIC), E-07122 Palma de Mallorca, Spain}
\author{David S\'{a}nchez}
\affiliation{Institute for Cross-Disciplinary Physics and Complex Systems IFISC (UIB-CSIC), E-07122 Palma de Mallorca, Spain}
\author{Michael Moskalets}
\affiliation{Institute for Cross-Disciplinary Physics and Complex Systems IFISC (UIB-CSIC), E-07122 Palma de Mallorca, Spain}
\affiliation{{Department~of~Metal~and~Semiconductor~Physics\char`,~NTU}~``Kharkiv Polytechnic Institute'', 61002~Kharkiv,~Ukraine}

\date{\today}

\begin{abstract}
 Identifying and controlling decoherence in single electron sources (SES) is important for their applications in quantum information processing. The recent experiments with ultrashort electron pulses [J. D. Fletcher {\it et al.,} Nat. Commun. 10, 5298 (2019)] demonstrate strong decoherence that cannot be caused by traditional mechanisms such as electron-electron or electron-phonon interactions. Here we propose timing noise as a universal model, consistent with existing experimental data, to explain strong decoherence of ultrafast SES pulses, without resorting  to any specific microscopic mechanism for such decoherence. We also propose a protocol to filter out timing noise 
which works even in the presence of other decoherence effects, such as those present in, e.g., low-energy SESs.
\end{abstract}

\maketitle

\section{Introduction}

Understanding the decoherence of electronic excitations is crucial for applying solid-state devices to quantum technologies.
In devices near equilibrium, decoherence is characterized by the coherence length.
This is determined by the interaction of the plane-wave electron at the Fermi-level with the surrounding electrons, \cite{Cabart:2018aa} phonons, and impurities~\cite{das_sarma_study_1980,taubert_relaxation_2011}.
In the case of interferometers such as Mach-Zehnder~\cite{ji_electronic_2003} or Fabry-P\'{e}rot~\cite{mcclure_edge-state_2009}, the competition between the coherence length and the size of the interferometer determines the visibility.

Decoherence is a significant problem in single-electron sources (SESs)~\cite{Splettstoesser:2017jd,bauerle_coherent_2018,pekola_single-electron_2013,keeling_minimal_2006,dubois_minimal-excitation_2013,feve_-demand_2007,moskalets_quantized_2008,keeling_coherent_2008,giblin_towards_2012,kaestner_non-adiabatic_2015,ryu_ultrafast_2016,hermelin_electrons_2011}
because these systems are typically very sensitive to external noise.
SESs generate nonequilibrium electron excitations by AC driving, in an on-demand fashion. This leads to time-resolved studies of the coherent oscillations~\cite{feve_-demand_2007,kataoka_coherent_2009,kataoka_tunable_2011,yamahata_picosecond_2019}, Fermionic statistics~\cite{bocquillon_electron_2012,bocquillon_coherence_2013,wahl_interactions_2014,grenier_fractionalization_2013}, and Coulomb interaction~\cite{ubbelohde_partitioning_2015,ryu_partition_2022,fletcher_time-resolved_2023,ubbelohde_two_2023,wang_coulomb-mediated_2023} of a few electron excitations.
These works suggest the possibility towards realizing electron flying qubits~\cite{bauerle_coherent_2018}.
In low-energy SESs, such as Leviton pumps~\cite{keeling_minimal_2006,dubois_minimal-excitation_2013} and mesoscopic capacitors~\cite{Gabelli:2006eg,feve_-demand_2007,moskalets_quantized_2008,keeling_coherent_2008}, the decoherence effect after the emission from the SESs is generally caused by interaction with the surrounding electrons~\cite{bocquillon_electron_2012,bocquillon_coherence_2013,wahl_interactions_2014,ferraro_real-time_2014} and impurities~\cite{acciai_influence_2022}.
Two-particle interferometer platforms, electronic analogue of the Hong-Ou-Mandel setup, show an incomplete shot-noise suppression, as opposed to the expectation by the Pauli-exclusion principle, which can be well explained by such decoherence~\cite{bocquillon_coherence_2013,wahl_interactions_2014,iyoda_dephasing_2014}.


On the other hand, in high-energy SESs, namely, quantum-dot pumps~\cite{Blumenthal:2007ho,kataoka_tunable_2011,giblin_towards_2012,fletcher_clock-controlled_2013, kaestner_non-adiabatic_2015,ryu_ultrafast_2016}, the decoherence after the emission is small because the single-electron excitation is effectively isolated from the Fermi sea and the phonons~\cite{taubert_relaxation_2011,emary_phonon_2016,johnson_lo-phonon_2018}.
However, recent tomography experiments reveal that the electron has quantum purity as low as $\sim$0.04~\cite{fletcher_continuous-variable_2019}.
This suggests that decoherence intrinsic to the SES is significant.
The time uncertainties in the excitations of the high-energy SES ($\sim 5 $ ps~\cite{fletcher_continuous-variable_2019}) are  much smaller than those of low-energy ones ($\sim 40 $ ps~\cite{jullien_quantum_2014}) hence a small classical fluctuation in timing  (e.g., caused by jitter in time-dependent voltage~\cite{fletcher_clock-controlled_2013}) might cause a severe decoherence in high-energy SESs.

\begin{figure}[b]
  \centering
  \includegraphics[width=\columnwidth]{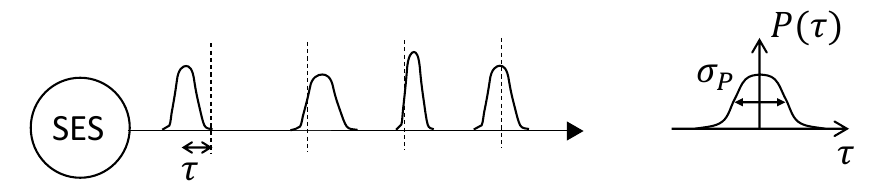}
  \caption{Single-electron sources with timing noise.
    At each emission (dashed lines), electron wave packet is temporally translated by random timing shift $\tau$, governed by probability distribution $P(\tau)$, possibly with additional distortions due to other  noise (noise to the packet shape is considered in the figure). }
  \label{fig:setup}
\end{figure}

In this work, we study the decoherence effect induced by timing noise in SESs, see Fig.~\ref{fig:setup}.
This noise generates a stochastic ensemble of wave packets which are temporally translated. Thus, the coherence between different energy components is suppressed when the time uncertainty of the noise is much larger than that of the wave packet. 
We present a protocol to identify the timing noise, i.e., to decide whether a SES involves timing noise and to obtain its noise distribution.
The protocol also enables to extract the time uncertainty of the timing noise that characterizes the noisy SES.
We show that an energy filtering can recover the coherence when the time elongation due to the filtering is larger than the time uncertainty of the timing noise.

This paper is organized as follows. In Sec.~\ref{sec:tNoise}, we develop the theory of the timing noise. In Sec.~\ref{sec:general-method}, we show the identification protocol and an example applied to the high-energy SES with timimg and shape noise. In Sec.~\ref{sec:simp-method}, we present a simpler method to identify the timing noise as the only source of decoherence. This method shows that the low-purity bivariate Wigner function, observed in Ref.~\cite{fletcher_continuous-variable_2019}, is consistent with the timing noise. In Sec.~\ref{sec:tNoiseCan}, the protocol to mitigate the timing-noise induced decoherence is presented. The conclusion is given in Sec.~\ref{sec:con}.

\section{Single-electron source with timing noise}
\label{sec:tNoise}
Periodically, the source generates a single-electron excitation, described by a density matrix $\chi$ of time uncertainty $\sigma_t$ (namely, the position uncertainty divided by the velocity), with random time shifts $\tau$ distributed according to the probability function $P(\tau)$ of variance $\sigma_P^2$, see Fig.~\ref{fig:setup}.
Note that the source generates the excitation in periodic fashion to generate an electric current sufficiently large enough to be measured. 
We assume that the period $T$ is much larger than the time uncertainties of the excitations to ensure stochastic independence between the periods, $T \gg \sigma_t , \sigma_P$.
We also assume that the waveguide into which the SES emits the excitation has a linear dispersion relation between energy and momentum to avoid the wave-packet spreading.
We assume that the noise distribution is an even function, $P(-\tau)=P(\tau)$, as the noise does not prefer a specific time direction. 

The random timing shift can be caused by various mechanisms, e.g., a jitter in arbitrary waveform generator~\cite{fletcher_clock-controlled_2013} or by charge fluctuations which affect the potential of the quantum dot, leading to emission delay~\cite{kataoka_time-resolved_2017}.
Rather than studying the microscopic mechanism, we focus on the characterization and identification of the state generated by the noisy SES.

The electron state generated by the noisy SES is given by the stochastic ensemble of wave packets 
$\chi$ 
with different timings, and hence described by the density matrix
\begin{equation}
\rho = \int d\tau \, P(\tau) \hat{\mathbf{T}}(\tau) 
  \chi \hat{\mathbf{T}}^\dagger(\tau)   
  .
  \label{eq:rho}
\end{equation}
Here $\hat{\mathbf{T}}(\tau)$ is the translation operator which delays the emission timing by $\tau$. In general, $\chi$ is a mixed state.
For example, a SES may involve a noise in a parameter determining the shape of the wave packet, which would result in the mixed-state $\chi$, see Sec.~\ref{sec:general-method}. When the timing noise is the only source of the decoherence, $\chi$ is a pure state.
In terms of Wigner distribution, Eq.~(\ref{eq:rho}) is equivalent to
\begin{equation}
  W(E,t) = \int d\tau\, P(\tau) W_0(E, t-\tau) .
  \label{eq:W}
\end{equation}
where $W$ is the Wigner distribution of the electron state $\rho$, and the connection between the Wigner distribution and the density matrix is
$
W(E,t) = (2/h) \int d\mathcal{E} \braket{E +\mathcal{E} |\rho |E -\mathcal{E}}
e^{-i 2\mathcal{E}t/\hbar} .
$
$W_0(E,t)$ is determined by the same equation replacing  $\rho$ with $\chi$.
Note that $t$ is the arrival time~\cite{emary_phonon_2016} which is an observable equivalent to position due to the linear dispersion relation.

Equation~(\ref{eq:rho}) provides the properties of the mixed state $\rho$ generated by the noisy SES.
Due to the convolution form of Eq.~(\ref{eq:W}), the temporal variance of the mixed state, $\Sigma_t^2\equiv \int dE \,dt\, t^2 W(E,t) - \{\int dE\, dt \, t W(E,t)\}^2$, is the summation of variances of the timing noise and the wave packet,
\begin{equation}
  \Sigma_t^2 = \sigma_P^2  + \sigma_t^2 ,  
\end{equation}
for any $P(\tau)$ and $\chi$.
The convolution form also suggests that the coherence stored in different energies is reduced.
In fact, the timing noise induces pure dephasing~\cite{Brouwer:1997aa,Buks:1998aa,Whitney:2008aa,mercurio_pure_2023} in the energy basis, which can be seen by Fourier transforming Eq.~(\ref{eq:W}),
\begin{equation}
  \braket{E|\rho|E'}
  = \braket{E| \chi |E'} \int d\tau \, P(\tau) e^{i (E-E')\tau/\hbar} .
  \label{rho-chi}
\end{equation}
The integral, i.e., the Fourier transform of $P(\tau)$, equals 1 for $E=E'$ [due to the normalization of $P(\tau)$] and decays when the energies $E$ and $E'$ differ more than $\hbar/\sigma_P$;
In the case of Gaussian noise, the integral becomes $e^{-\sigma_P^2 (E-E')^2/(2 \hbar^2)}$.
It follows that timing noise suppresses the off-diagonal elements while keeping the diagonal elements the same, $\braket{\rho}_E = \braket{\chi}_E$.
The coherence stored in different energies is almost lost when the variance of the timing noise is much larger than the time uncertainty of $\chi$, that is, \( \sigma_P \gg \sigma_t \). This is clearly seen for a Gaussian noise and a Gaussian packet $\chi$, when the quantum purity, \(\gamma = \text{Tr} \rho^2\), of the mixed state is determined as $\gamma = 1/\sqrt{1+ (\sigma_P/\sigma_t)^2}$.

\section{Identifying the timing noise}
\label{sec:iden}
We put forward a method that not only tells us if a SES involves timing noise or not but also provides details about the noise distribution $P(\tau)$ and the original state $\chi$.
Such identification is not a trivial task because the microscopic mechanisms are difficult to be controlled, and neither the noise distribution $P(\tau)$ nor $\chi$ are known in advance. 
Our protocol only relies on the information about the density matrix of $\rho$ which can be obtained from tomography, see ~\cite{fletcher_continuous-variable_2019}.

\begin{figure*}[t]
    \centering
    \includegraphics[width=0.8\textwidth]{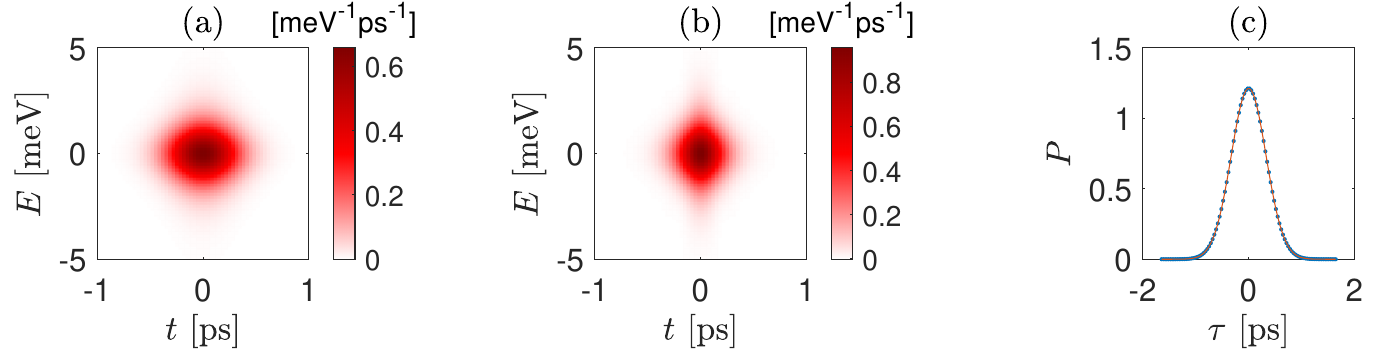} 
    \caption{Detecting timing noise for high-energy SES with timing and shape noise.
    (a) Wigner function of $\rho$ given by Eqs.~\eqref{eq:rho},\eqref{eq:chi-sNoise}, and Gaussian timing noise. 
    (b) Wigner function of the numerical solution of Problem~\ref{sdp} applied to $\rho$. 
    (c) The timing noise distribution numerically obtained from (a), (b), and Eq.~\eqref{eq:tP} (dots), and Gaussian fit (line) of standard deviation of 0.33 ps.
    Parameters for $\rho$: $w_0=0.33$ ps, $\sigma_w= 0.3 w_0$, $\sigma_P = 0.33$ ps, and the energy $E$ is measured from $E_0$}
    \label{fig:sdp_hotSES}
\end{figure*}

\subsection{General method }
\label{sec:general-method}
We find that the timing noise effect can be detected by solving the following optimization problem. 
\begin{problem}
For a given $\rho$, 
let $\tilde{\chi}$ be a positive-semidefinite matrix having the same dimension as $\rho$, 
and let a function $\tilde{g}(\mathcal{E})$ and its Fourier transform $\tilde{P}(\tau)$ be 
\begin{align}
    \tilde{g}(\mathcal{E})
    & \equiv 
    \frac{\braket{\bar{E} + \mathcal{E}/2 |\rho|\bar{E}-\mathcal{E}/2}}{\braket{\bar{E} +\mathcal{E}/2|\tilde{\chi}|\bar{E} -\mathcal{E}/2}} ,\label{eq:tg0}\\
    \tilde{P}(\tau)
    &\equiv 
    \frac{1}{h} \int d\mathcal{E} \, \tilde{g}(\mathcal{E})
    e^{-i \mathcal{E}\tau /\hbar} ,
    \label{eq:tP}
\end{align}
where $\bar{E}\equiv \int dE \, E \braket{E|\rho|E}$ is the mean energy of $\rho$.
Find $\tilde{\chi}$ which maximizes the variance 
$\sigma_{\tilde{P}}^2 \equiv \int d\tau \, \tau^2 \tilde{P}(\tau) $ while satisfying the constraints: 
\begin{align}
    &\frac{\partial}{\partial E} \frac{\braket{E+\mathcal{E}| \tilde{\chi}  |E-\mathcal{E}} }{\braket{E+\mathcal{E}| \rho|E-\mathcal{E}}}
     =0 , \label{eq:Opt-const1}
     \\
    &\text{Im}\, \big[\braket{E|\tilde{\chi}|E'} \braket{E'|\rho|E} \big]
    =0 , \label{eq:Opt-const2}
    \\
    &\braket{E|\tilde{\chi}|E} =\braket{E|\rho|E} . \label{eq:Opt-const3}
    \\
    & \tilde{P}(\tau) \ge 0 . 
    \label{eq:Opt-const4}
\end{align}
\label{sdp}
\end{problem}
Problem \ref{sdp} is equivalent to extracting the timing noise maximally from $\rho$. 
See Appendix \ref{sec:opt} for the detailed derivation and numerical method to solve the problem.
The solution of $\tilde{\chi}$ and corresponding $\tilde{P}(\tau)$ [determined by Eqs.~\eqref{eq:tg0} and \eqref{eq:tP}] are $\chi$ and $P(\tau)$, respectively.

Here we show an example using Problem~\ref{sdp} for detecting the timing noise in high-energy SES with timing and shape noise,
i.e., where the emitted wave packet at each emission is not only shifted with a random timing but also distorted with a random shape change. 
We consider a simple type of the shape noise, where only the width of the emitted wave packet at each emission is randomly changed, see Fig.~\ref{fig:setup}.  
Namely, the density matrix for the electron in the absence of the timing noise is described as, 
\begin{align}
    \chi
    &= \int d w\, Q(w)  \ket{\psi_w}\bra{\psi_w} , \label{eq:chi-sNoise}\\
    \braket{E|\psi_w} 
    &= \left(\frac{2w^2}{\pi\hbar^2} \right)^{1/4} \exp\Big[-\frac{w^2 (E-E_0)^2}{\hbar^2} \Big] , \\
    Q(w)
    &= \frac{1}{\sqrt{2 \pi \sigma_w^2}} \exp\Big[- \frac{(w-w_0)^2}{2\sigma_w^2} \Big]  . \label{eq:sNoise}
\end{align}
Here $\ket{\psi_w}$ is the Gaussian packet with mean energy $E_0$ and time uncertainty $w$. 
$Q(w)$ is the probability distribution for emitting the packet $\ket{\psi_w}$, which is assumed to be Gaussian with mean value $w_0$ and variance $\sigma_w^2$. 
In the presence of the timing noise, the density matrix of the electron is described by Eqs.~\eqref{eq:rho} and \eqref{eq:chi-sNoise}.

Figure~\ref{fig:sdp_hotSES}
shows the result of solving Problem~\ref{sdp} for an input state $\rho$ given by Eqs.~\eqref{eq:rho} and \eqref{eq:chi-sNoise} with a Gaussian $P(\tau)$. 
To numerically solve the problem, we used CVX, a package for specifying and solving convex programs~\cite{cvx,gb08}. 
The solution correctly detects Eq.~\eqref{eq:chi-sNoise} within trace distance of $10^{-5}$.
The numerically obtained timing noise distribution also correctly detects the Gaussian timing noise.
Note that in this result, the purities are 
$\text{Tr}\, \chi^2 = 0.91$, $\text{Tr}\, \rho^2 = 0.65$.

Note that our theory of the timing noise and the identification method also applies to low-energy SESs, although the effect of timing noise may be weak compared to other decoherence effects. Historically low-energy SESs are characterized by the excess-first order coherence~\cite{Grenier:2011dv} rather than by a density matrix. Then the effect of timing noise is taken into account by Eq.~\eqref{eq:rho}, where the single-electron density matrices $\rho$ and $\chi$ are substituted by the corresponding coherences, see Eq.~\eqref{eq:C-c} and Appendix~\ref{sec:coldSES}.

\subsection{Timing noise as the only source of decoherence}
\label{sec:simp-method}
We show a simpler method which can identify the timing noise as the only source of the decoherence. 
For simplicity, we assume that the density matrix $\braket{E|\rho|E'}$ does not have zeros, see Appendix \ref{sec:apd_theorems} for the contrary case.
The method is based on the following theorem.

\textbf{Theorem:}
A mixed state $\rho$ satisfies Eq.~(\ref{eq:rho})  with pure state $\chi$ if and only if (i) the function 
$
 |\braket{E|\rho|E'}|/\sqrt{\braket{\rho}_E \braket{\rho}_{E'}} 
$
only depends on energy difference $E-E'$ and (ii) its Fourier transform with respect to $E-E'$ is a positive function, and (iii) $e^{i \text{ arg} \braket{E|\rho|E'} }$ is a rank-1 matrix.
We refer the reader to Appendix~\ref{sec:simple} for the proof. 
When these conditions hold, the noise distribution $P(\tau)$  and the wave packet $\chi = \ket{\psi}\bra{\psi}$ are uniquely determined in terms of the density matrix as
\begin{align}
    &P(\tau)
  = \frac{1}{h}\int d E
    \frac{|\braket{\bar{E} +\frac{E}{2} |\rho|\bar{E} -\frac{E}{2}}|}
    {\sqrt{\braket{\rho}_{\bar{E} +\frac{E}{2}} \braket{\rho}_{\bar{E} -\frac{E}{2}}}}
    e^{-i E\tau/\hbar} ,
    \label{eq:P-rho} \\
  &\braket{E|\psi}\braket{\psi |E'}
  = \sqrt{\braket{\rho}_{E} \braket{\rho}_{E'} } e^{i \text{ arg} \braket{E|\rho|E'}} ,
  \label{eq:chi-rho}
\end{align}
where $\bar{E} \equiv \int dE\, E \braket{\rho}_E $ is the mean energy.
Note that the noise distribution in Eq.~\eqref{eq:P-rho} is the Fourier transform of the function used in the condition (i).

Using the identification method, we find that a mixed state whose Wigner distribution is bivariate Gaussian, experimentally detected in quantum-dot SESs~\cite{fletcher_continuous-variable_2019}, is consistent with timing noise.
The Wigner distribution is written as
\begin{equation}
W(E,t) =  
\frac{
\exp\left[ -  \frac{1}{2(1-r)^2} \left(  \frac{E^{2}}{\Sigma_E^{2}} -2r \frac{E}{\Sigma_E}\frac{t}{\Sigma_t} +\frac{t^{2}}{\Sigma_t^{2}}\right) \right]
}{2\pi \Sigma_E \Sigma_t \sqrt{1-r^2}} ,
\label{eq:wbg}
\end{equation}
where $\Sigma_E$, $\Sigma_t$, and $r$ are the energy, time uncertainties and the energy-time correlation, respectively.
These states are observed in Ref.~\cite{fletcher_continuous-variable_2019} in the regime of fast pumping.
One can easily verify the conditions (i)--(iii).
Using Eqs.~(\ref{eq:P-rho}) and (\ref{eq:chi-rho}), we obtain both the noise distribution and the wave packet,
\begin{align}
    P(\tau)
  &= \frac{1}{\sqrt{2\pi} \sigma_P } e^{- \tau^2/(2\sigma_P^2)} , \label{eq:P-biG}\\
  \sigma_P
  &=  \sqrt{  (1-r^2) \Sigma_t^2 -\big(\frac{\hbar }{2 \Sigma_E}\big)^2 } ,
    \label{eq:sigma-biG} \\
  \braket{E|\psi}
  &= \frac{1}{\sqrt{\sqrt{2\pi} \Sigma_E}}
    \exp \Big[ -\frac{E^2}{4 \Sigma_E^2}  +i \frac{r \Sigma_t E^2}{2\hbar \Sigma_E} \Big] . \label{eq:psi-biG}
\end{align}
Note $\sigma_P$ in Eq.~(\ref{eq:sigma-biG}) is manifestly real due to the fact that the purity is 
bounded by \( 1 \),
$  \hbar/[2\sqrt{1-r^2} \Sigma_t \Sigma_E] \le 1$.
\begin{figure}[t]
  \centering
  \includegraphics[width=\columnwidth]{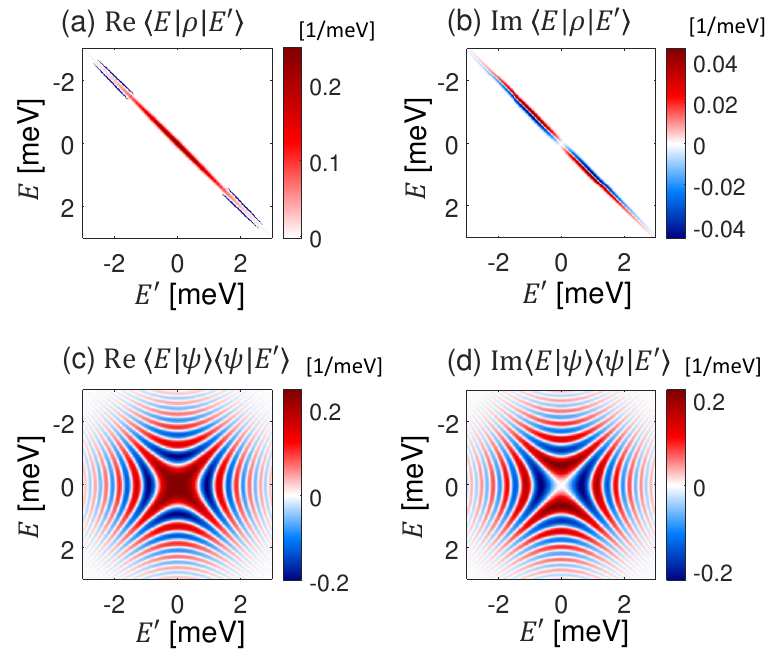}
  \caption{Identification of the timing noise as the only source of decoherence.
    (a)--(b) The density matrix in energy basis for the state detected in Ref.~\cite{fletcher_continuous-variable_2019}. Its Wigner distribution is bivariate Gaussian with $\Sigma_E=1.61$ meV, $\Sigma_t=6.44$ ps, and $r=0.47$.
    This state is the result of noisy SES emitting pure-state wave packet $\psi$, (c)--(d), with Gaussian timing noise of uncertainty $\sigma_P = 5.7 $ ps. }
  \label{fig:rho}
\end{figure}

Fig.~\ref{fig:rho} shows the density matrix of $\rho$ and its wave-packet component. 
It shows that the mixed state, being squeezed along the diagonal in the energy basis [see panels (a) and (b)]  is significantly different from the wave-packet component, which exhibits high degree of symmetry [see panels (c) and (d)].

It is worth mentioning a peculiarity of bivariate Gaussian Wigner distribution. On one hand, it is consistent with timing noise. On the other hand, because the energy and time appear in equal footing in such distribution, it can also be interpreted by another description of an energy noise. We find that a bivariate Gaussian can be written as 
$W(E,t) = \int d\mathcal{E} \, \tilde{P}(\mathcal{E}) \tilde{W}_0 (E -\mathcal{E}, t) $ where $\tilde{P}(\mathcal{E})$ is a Gaussian noise distribution for random energy shifts applied to a pure-state bivariate Gaussian $\tilde{W}_0(E,t)$, see Appendix~\ref{sec:eNoise}.
Note that such alternative interpretation is generally impossible for non-Gaussian states; e.g., a state generated by a (zero-temperature) Leviton source with a timing noise can not be explained by another source with the energy noise.
Furthermore, the fact that a bivariate Gaussian state is consistent with timing noise allows for a simple and intuitive purification procedure as follows.

\section{Cancelling timing noise}
\label{sec:tNoiseCan}
Luckily, the coherence loss due to timing noise can be recovered.
The purity reduction is determined by the competition of noise time uncertainty $\sigma_P$ and the quantum time uncertainty $\sigma_t$. 
Consider a dynamic mapping which changes the packet $\psi$ to another pure state $\psi'$ which is elongated in time, so that the new time uncertainty $\sigma_t'$ is larger than the original value  $\sigma_t$.
If the dynamic mapping commutes with the time-translation operator $\hat{\mathbf{T}}(\tau)$, the mapping changes the mixed state $\rho$ to
\begin{equation}
  \rho' = \int d \tau\, P(\tau) \hat{\mathbf{T}}(\tau) \ket{\psi'} \bra{\psi'}
  \hat{\mathbf{T}}^\dagger (\tau) .
  \label{eq:rhop}
\end{equation}
Then, the new purity $\gamma' = \text{Tr} (\rho')^2$ becomes larger than the original purity $\gamma = \text{Tr} \rho^2$ since the relative strength of the timing noise is weaker, namely, $\sigma_P/ \sigma_t' < \sigma_P/\sigma_t$. 
Note that  $\sigma_t'$ should be smaller than the SES period to ensure the interperiod independence.

\begin{figure}[b]
  \centering
  \includegraphics[width=\columnwidth]{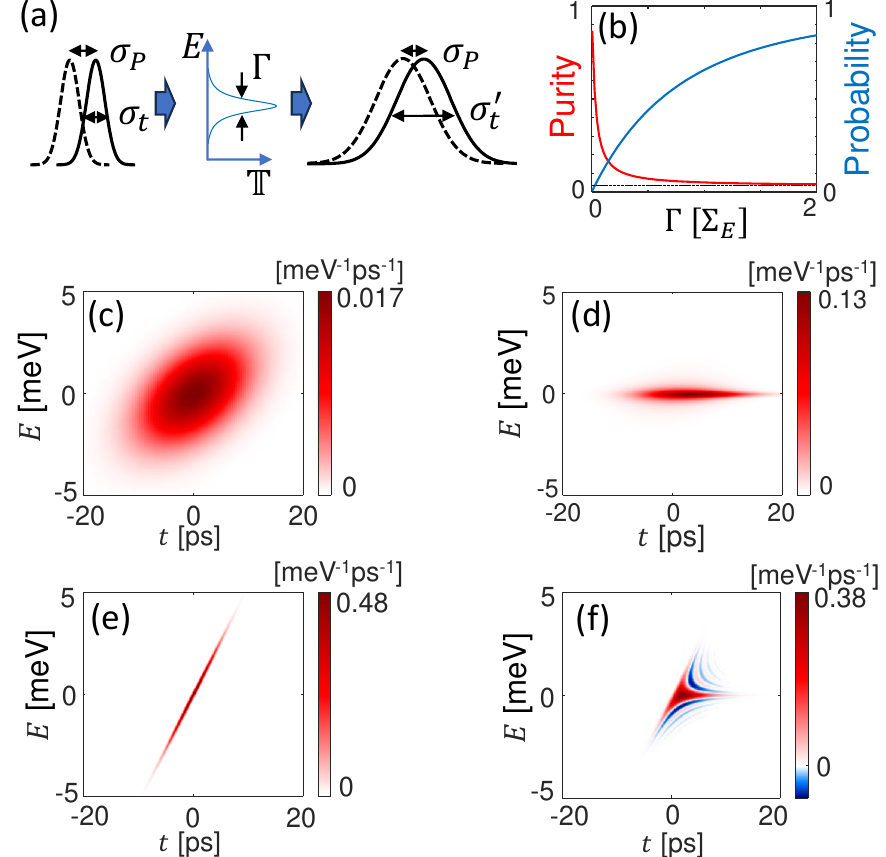}
  \caption{(a) Scheme for the timing noise cancellation using a resonance level.
    A mixed state generated by temporally noisy SES is energy filtered by a resonance level with broadening $\Gamma$ and becomes elongated in time, reducing the effect of the timing noise.
    (b) Purity after the filtering $\gamma'$ and the packet transmission probability $\mathbb{T}_0$. Dotted line: purity before the filtering $\gamma$.
    (c)--(d) Wigner distribution before and after the filtering at $\Gamma=0.1 \Sigma_E$. The purity is increased from 0.036 to 0.22 with $12\%$ probability.
    (e)--(f) Wigner distribution of the pure-state components, $\psi$ and $\psi'$, corresponding to (c) and (d).
    Note that here we show the Wigner distribution instead of the density matrix for a compact visualization but they are equivalent; the panels (c) and (e) are equivalent to Fig.~\ref{fig:rho}. Parameters: the state of Fig.~\ref{fig:rho} is used for the state before the filtering. }
  \label{fig:nCancel_1lev}
\end{figure}

\begin{figure}[t]
  \centering
  \includegraphics[width=\columnwidth]{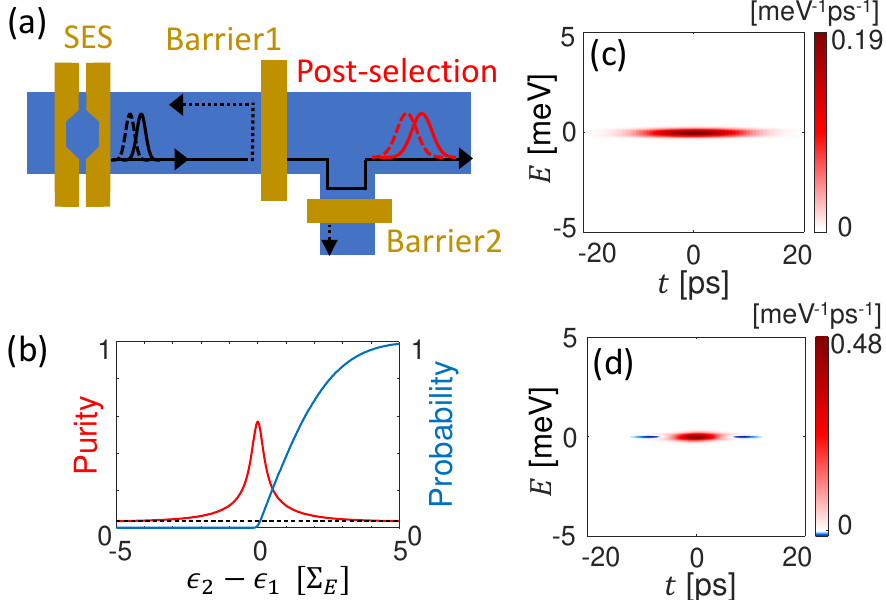}
  \caption{(a) 
    Energy filtering using potential barriers without backscattering between them.
    The state which has transmitted through the first barrier and reflected at the second one is post-selected.
    (b) Enhanced purity and the post-selection probability for various barrier height differences $\epsilon_2-\epsilon_1$. Dashed line: purity before the filtering $\gamma$.
    (c)--(d) Wigner distribution after the filtering and its pure-state component at $\epsilon_2-\epsilon_1 =0.2 \Sigma_E$ where the purity is enhanced from 0.036 to 0.41 with $8\%$ probability. 
    Parameter: $\Delta_b = 0.03 \Sigma_E$ and the state of Fig.~\ref{fig:rho} is used for the state before the filtering.}
  \label{fig:nCancel_TR}
\end{figure}

Such dynamic mapping should be nonunitary because a unitary operation conserves the purity. 
Additionally, $\psi'$ should have smaller energy uncertainty than $\psi$ in order to have larger time uncertainty. 
Therefore, a natural option for the dynamic mapping is energy filtering.
This energy filtering can be realized by postselecting a state transmitted through a single resonance level with a large lifetime (i.e., small resonance width $\Gamma$), see Fig.~\ref{fig:nCancel_1lev}(a).
The state post-selected by the energy filtering is described by Eq.~(\ref{eq:rhop}) with  
\begin{equation}
  \ket{\psi'} = \frac{1}{\sqrt{\mathbb{T}_0}}
  \int dE\, \ket{E}\braket{E|\psi} \mathbb{t}(E) 
  \label{eq:psip}
\end{equation}
where $\mathbb{t}(E)$ is the transmission amplitude, $\mathbb{T}(E) \equiv |\mathbb{t}(E)|^2$ is the corresponding transmission probability, and $\mathbb{T}_0 = \int dE \, |\braket{E|\psi}|^2 \mathbb{T}(E) $ is the transmission probability for the packet.

In Fig.~\ref{fig:nCancel_1lev} we show the result of calculations using the transmission probability $\mathbb{T}(E) = 1/[1+ E^2/\Gamma^2] $, where energy is measured from the resonance level. 
For simplicity, we focus on the case that the mean energy of $\rho$ is aligned with the resonance level (see Appendix~\ref{sec:ext_nCancel} for different cases).
Panel (b) in Fig.~\ref{fig:nCancel_1lev} 
shows the purity $\gamma'$ after filtering a bivariate-Gaussian Wigner distribution as observed in Ref.~\cite{fletcher_continuous-variable_2019}.
The purity approaches 1 as the level broadening $\Gamma$ is reduced.
However, a sharp filtering accompanies the weaker signal, manifested in the trade-off relation shown in the transmission probability and the state deformation. 
One should select the value of $\Gamma$ with a compromise.
For example, when $\Gamma=0.1 \Sigma_E$, the purity is significantly enhanced from  $\sim 0.04$ to $\sim 0.2$ with $10 \%$ probability, without completely deforming the state [see Fig.~\ref{fig:nCancel_1lev}(c)-(d)]. 
Note that the situation of single-resonance level can be implemented using a Fabry-P\'{e}rot interferometer (or a quantum dot) with a resonance-level spacing much larger than the energy uncertainty of the state before filtering. 

The Fabry-P\'{e}rot interferometer often involves complications from Coulomb interactions~\cite{rosenow_influence_2007,ofek_role_2010, halperin_theory_2011,yang_fabry-perot_2020}.
Instead, energy filtering can also be realized by a three-lead junction with two potential barriers without backscattering between them, readily implementable in the experiments, see Fig.~\ref{fig:nCancel_TR}(a).
Here the state after the transmission and subsequent reflection is post-selected.
The transmission probability of the first [second] barrier is 
$1/(1+\exp\{-(E-\epsilon_{1[2]})/\Delta_b\})$ \cite{fertig_transmission_1987}.
$\epsilon_{n=1,2}$ is the height of the $n$th barrier and $\Delta_b$ is the energy scale characterizing sharpness of the barriers.
We assume $\Delta_b \ll \Sigma_E$ to have sufficient filtering effect and ignore the small phase variation~\cite{barratt_asymmetric_2021} in the tunneling amplitude over the energy window $E \in [\epsilon_n-\Delta_b, \epsilon_n +\Delta_b]$.
Fig.~\ref{fig:nCancel_TR}(b)--(d) shows the result, focusing on the case that $\Delta_b = 0.03\Sigma_E$ and $(\epsilon_1+\epsilon_2)/2 = \braket{E}_\rho $; see Appendix~\ref{sec:ext_nCancel} for other cases.
Similarly to the result of Fig.~\ref{fig:nCancel_1lev}, we obtain a significant purity enhancement with sufficient probability, e.g., the purity is enhanced from $\sim 0.04$ to $\sim$ 0.4 with $8 \%$ probability at $\Delta_b = 0.03\Sigma_E$ and $\epsilon_2-\epsilon_1 = 0.2\Sigma_E $.
The detailed form of the post-selected packet $\ket{\psi'}$ shown in Fig.~\ref{fig:nCancel_TR}(d) differs from that of Fig.~\ref{fig:nCancel_1lev}(f) due to the different forms of the transmission amplitude $\mathbb{t}(E)$ (rectangular function for the former and simple pole for the latter).

\section{Conclusion}
\label{sec:con}
We have developed a theory for timing noise present in single-electron sources. Timing noise induces a pure dephasing in energy basis. 
This fact provides a method for identifying SES with such noise, extracting information about the noise, and allows for a simple and experimentally accessible purification procedure based on energy filtering. 
In general, timing noise can be expected to become increasingly relevant as the voltage pulse durations approach
the intrinsic time scale of the source.
Thus, our theory appears to be relevant not only to single-electron sources of type used in Ref.~\cite{fletcher_continuous-variable_2019}, but also to the emerging fields of research requiring ultrafast  voltage pulses~\cite{Georgiou2020,Aluffi_2023}.

\section{Acknowledgments}
This work was partially supported by the Spanish State Research Agency (MCIN/AEI/10.13039/501100011033) and FEDER (UE) under Grant
No.\ PID2020-117347GB-I00 and the Mar{\'\i}a de Maeztu project CEX2021-001164-M,
and by CSIC/IUCRAN2022 under Grant No.\ UCRAN20029. SR acknowledges a partial support from UIB through the research project led by a postdoctoral principal investigator.

\appendix

\section{Optimization problem for identifying timing noise}
\label{sec:opt}

Here we show the proof that solving Problem~\ref{sdp} is equivalent to identify the timing noise, i.e., finding $\chi$ and $P(\tau)$ for given $\rho$. We also explain how the problem can be solved numerically.

We recall that a state affected by the timing noise is described by Eq.~(1), 
\begin{equation}
    \rho = \int d \tau \, P(\tau) \hat{\mathbf{T}}(\tau) \chi \hat{\mathbf{T}}^\dagger(\tau) .
    \label{eq:rho-chi}
\end{equation}
In energy basis, Eq.~\eqref{eq:rho-chi} is equivalently written as
\begin{equation}
    \braket{E|\rho|E'} = \braket{E| \chi| E'} g(E-E') ,
\end{equation}
where $g(\mathcal{E}) \equiv \int d \tau \, P(\tau) e^{i \mathcal{E} \tau/\hbar}$ is the Fourier transform of the timing noise distrtibution. 


The constraint~\eqref{eq:Opt-const1} of Problem~\ref{sdp} is equivalent to $\braket{E|\rho|E'} = \braket{E|\tilde{\chi}|E'}\tilde{g}(E-E')$, hence equivalent to
\begin{equation}
\rho = \int d \tau \, \tilde{P}(\tau) \hat{\mathbf{T}}(\tau) \tilde{\chi} \hat{\mathbf{T}}^\dagger(\tau) .    
\label{eq:rho-X}
\end{equation}
The constraint~\eqref{eq:Opt-const2} is equivalent to that $\tilde{g}(\mathcal{E}) $ is real and even function, hence $\tilde{P}(\tau)$ is real and even function.
The constraint~\eqref{eq:Opt-const3} is equivalent to that $\tilde{g}(0)=1$, hence $\int d\tau\, \tilde{P}(\tau) =1$.
Hence, the constraints~\eqref{eq:Opt-const2}--\eqref{eq:Opt-const4} ensure that $\tilde{P}(\tau)$ is a physical timing noise distribution.
There exist various $\tilde{\chi}$ and $\tilde{P}$ depending on the level of the timing noise extraction, from the minimal extraction, $\tilde{\chi}=\rho$, $\tilde{P}(\tau)=\delta(\tau)$, to the maximal extraction $\tilde{\chi}=\chi$, $\tilde{P}(\tau) = P(\tau)$.
Finding $\tilde{\chi}$ with maximum $\sigma_{\tilde{P}}^2$ corresponds to extract the timing noise maximally. 
Note that when $\rho$ does not involve the timing noise, the  trivial solution is obtained, i.e., $\tilde{\chi}$ = $\rho$ and $\tilde{P}(\tau) = \delta(\tau)$.

One can simplify the optimization target of Problem~\ref{sdp} which is beneficial for numerical optimization.
Maximizing $\sigma_{\tilde{P}}^2$ is equivalent to minimizing the arrival-time variance of the extracted state, 
namely $\text{Tr} (\hat{t}^2 \tilde{\chi}) -(\text{Tr} \,\hat{t}\tilde{\chi})^2$ where $\hat{t}$ is the observable for the arrival time.
This is because that the sum of the two variances is the arrival-time variance of $\rho$ which is constant during the optimization. 
The second term in the minimization target, $ -(\text{Tr} \,\hat{t}\tilde{\chi})^2$, is also a constant because $ \text{Tr} \,\hat{t}\tilde{\chi}= \text{Tr}\, \hat{t}\rho$. 
The equality holds because of the symmetry of the timing noise, $\int d\tau \, \tau \tilde{P}(\tau) =0$; 
One can easily check
$\text{Tr} \, \hat{t} \rho = \int d\tau \, dt \, t \tilde{P}(\tau) \braket{t-\tau |\tilde{\chi}| t-\tau} =  \int d\tau \, dt' \,  (t'+\tau) \tilde{P}(\tau) \braket{t'|\tilde{\chi}| t'}
= \int d\tau \, dt' \,  t'\tilde{P}(\tau) \braket{t'|\tilde{\chi}| t'}
= \text{Tr} \, \hat{t}\tilde{\chi}$,
where $\ket{t}$ is the eigenstate of arrival time (or equivalently position due to the linear dispersion relation between the energy and momentum),
the change of variable $t' \equiv t-\tau$ is used in the second equality, 
and $\int d\tau \, \tau \tilde{P}(\tau) = 0$ is used in the third equality. 
Hence, the optimization target of Problem~\ref{sdp} is simplified to minimizing $\text{Tr} (\hat{t}^2 \tilde{\chi})$.

Problem~\ref{sdp} can be solved numerically and efficiently using a semidefinite programming~\cite{watrous_theory_2018,nocedal_numerical_2006}.
First, $\rho$ and $\tilde{\chi}$ are described by the density matrix in energy basis, discretized with small spacing $\Delta$. 
As explained above, the optimization target is simply minimizing $\text{Tr} (\hat{t}^2 \tilde{\chi} )$.
Hence the optimization target is a linear function in the optimization object $\tilde{\chi}$. 
The observable $\hat{t}^2$ is numerically described as $\braket{E|\hat{t}^2|E'} = -(\hbar/\Delta)^2  (\delta_{E,E'+\Delta}+\delta_{E,E'-\Delta} -2\delta_{E,E'})$. 
The constraints~\eqref{eq:Opt-const1}--\eqref{eq:Opt-const3} are also linear in $\tilde{\chi}$.
Hence, Problem~\ref{sdp} without the constraint~\eqref{eq:Opt-const4} is a semidefinite programming problem which can be solved numerically and efficiently.
When the solution of this problem gives unphysical timing noise distribution, namely $\tilde{P}(\tau)$ with some negative values, one can solve the original problem with all the constraints.
However, we find that the simplified problem with only the constraints \eqref{eq:Opt-const1}--\eqref{eq:Opt-const3} always give physical solution in realistic situations relevant to the single-electron sources such as those of Fig.~\ref{fig:sdp_hotSES} and Fig.~\ref{fig:sdp_coldSES}.


\section{Low-energy SES with timing noise}
\label{sec:coldSES}
Here we show that our theory about the timing noise, Sec.~\ref{sec:tNoise}, and the identification protocol, Sec.~\ref{sec:general-method}, are applicable to the low-energy SESs.

\begin{figure*}[t]
    \centering
    \includegraphics[width=0.85\textwidth]{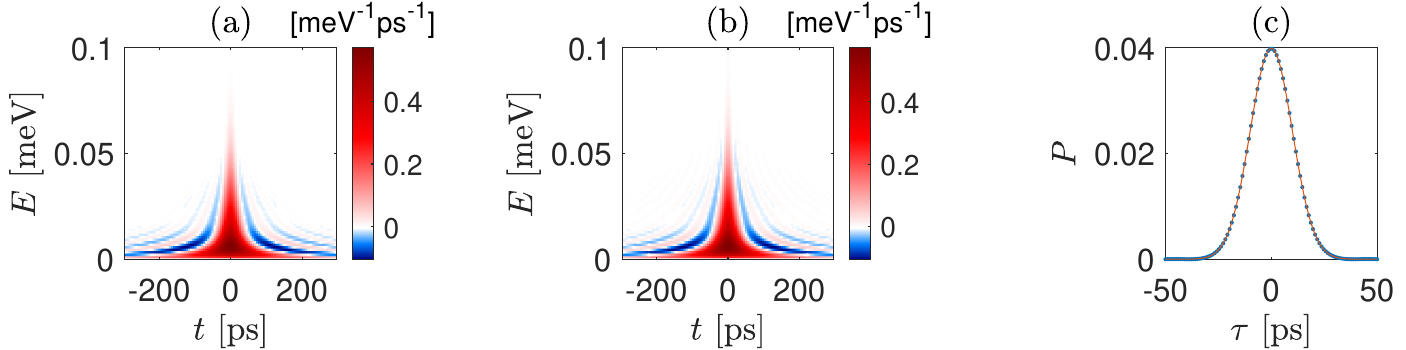} 
    \caption{Detecting timing noise for low-energy SES with timing noise.
    (a) Wigner function of $C$ given by Eqs.~\eqref{eq:C-c},\eqref{eq:c-Lev}, and Gaussian timing noise. 
    (b) Wigner function of the numerical solution of Problem~\ref{sdp} applied to $C$. 
    (c) The timing noise distribution numerically obtained from (a),(b) and Eq.~\eqref{eq:tP} (dots), and Gaussian fit (line) of standard deviation of 10 ps.
    Parameters for $C$: $w=42$ ps, $p_1=0.8$, $p_2 =0.2$ (corresponding to Leviton with temperature of 50 mK~\cite{bisognin_quantum_2019}), and $\sigma_P = 10$ ps.}
    \label{fig:sdp_coldSES}
\end{figure*}

In the case of low-energy electron sources, the timing noise effect is the same when substituting the single-electron density matrices to the excess first-order coherences~\cite{Grenier:2011dv},
\begin{equation}
    C(t,t') = \int d \tau \,  P(\tau) \, c(t-\tau, \, t'-\tau)  .
    \label{eq:C-c}
\end{equation}
Here $C$ and $c$ are the excess first-order coherences in the electronic subspace (i.e., the energy subspace above the Fermi-level) which corresponds to $\rho$ and $\chi$, respectively.
The only difference from the the single-electron density matrix is that the  first-order coherences are not necesarilly normalized; 
$\int dt\, C(t,t)$ and $\int dt\, c(t,t)$ are the numbers of electron excitations generated by the SES and can be deviated from 1 due to experimental imperfections.
Problem~\ref{sdp} formulated in terms of $C$ and $c$ works also in the case of a multi-particle emission.
Note that when treating $C$ and $c$ as operators~\cite{roussel_processing_2021}, e.g., defining $\hat{C} = \int dt \, dt' \,  C(t,t')\ket{t}\bra{t'}$, they are positive-semidefinite. 

The low-energy SESs generate the electron excitations near the Fermi-level, hence the interaction between the excitation and surrounding electrons/phonons becomes important. 
Here we consider a finite-temperature Leviton with timing noise.
First, we assume that there is no any timing noise. We consider the excitations generated by the Lorentzian voltage pulse with unit flux over the Fermi sea with subkelvin temperature. As it was demonstrated in Ref.~\cite{bisognin_quantum_2019}, these excitations are well described by the set of the two wave functions, \(\varphi_1\) and \(\varphi_2\). The corresponding excess first-order coherence in the electronic subspace is the following,
\begin{align}
   c(t,t') &=  p_1 \varphi_1(t) \varphi_1^*(t')
   +p_2 \varphi_2(t) \varphi_2^*(t') , 
   \label{eq:c-Lev}\\
   \braket{E|\varphi_1}&=
   \sqrt{\frac{2w}{\hbar}} e^{-E w/\hbar} \, \theta(E) , \\
   \braket{E|\varphi_2}&=
   \sqrt{\frac{2w}{\hbar}} e^{-E w/\hbar} \left(1- 2\frac{Ew}{\hbar}\right)  \theta(E) .
\end{align}
$\theta(E)$ is the Heaviside step function and the energies are measured with respect to the Fermi-level.
$\varphi_{n=1,2}$ is the wavepacket for the two dominant electron excitations which occurs with probability $p_{n}$ and $w$ is the half width at half maximum of the Lorentzian voltage pulse.
In ideal situation $p_1+p_2 =1$, but in experiments the equality can be slighly deviated due to the imperfections in the control of the voltage pulses. 
In the presence of the timing noise, the excess coherence is described by Eqs.~\eqref{eq:C-c} and \eqref{eq:c-Lev}.
Our protocol using Problem~\ref{sdp} works even when the equality is violated.

Fig.~\ref{fig:sdp_coldSES}
shows the result of solving Problem 1 for a coherence $C$ given by Eqs.~\eqref{eq:C-c},\eqref{eq:c-Lev}, and Gaussian timing noise. 
Again CVX package is used~\cite{cvx}. 
The solution correctly detects Eq.~\eqref{eq:c-Lev} within trace distance of $10^{-10}$.
The numerically obtained timing noise distribution also correctly detects the Gaussian timing noise.
The similarity between the Wigner functions of (a) and (b) demonstrates that our timing noise detection method is useful even for the regimes of weak timing noise.
Note that the quantum purity is 0.6811 and 0.6683 in the absence and presence of the timing noise, respectively.

\section{Theorem for identifying timing noise as the only source of decoherence}
\label{sec:apd_theorems}
We present the proof of the theorem in Sec.~\ref{sec:simp-method} used for the identification of the timing noise as the only source of decoherence.

\subsection{In the absence of zero of $\braket{E|\rho|E'}$}
\label{sec:simple}

As assumed in the main text, we first consider the simple case that the density matrix $\braket{E|\rho|E'}$ does not have any zero.

\textbf{Theorem:}
If there is no zero of $\braket{E|\rho|E'}$, Eq.~(1) with pure state $\chi$ holds if and only if (i) the function 
$ |\braket{E|\rho|E'}|/\sqrt{\braket{\rho}_E \braket{\rho}_{E'}} $
only depends on the energy difference $E-E'$ and (ii) its dependence has a positive Fourier transform, and (iii) $e^{i \text{arg} \braket{E|\rho|E'}}$ is rank-1 matrix.

{\it Proof.}
We first show that the conditions (i)--(iii) are necesary for Eq.~\eqref{eq:rho} with pure state $\chi$.
Using $\hat{\mathbf{T}}(\tau) \ket{E} = e^{i E \tau/\hbar} \ket{E}$,
Eq.~\eqref{eq:rho} is equivalent to 
\begin{equation}
  \braket{E|\rho|E'} = g(E-E') \braket{E|\chi|E'} ,
  \label{eq:timing-noise}
\end{equation}
where $g(E)\equiv \int d\tau P(\tau) e^{i E\tau/\hbar} $ is the Fourier transform of the timing noise distribution $P(\tau)$ and $\chi \equiv \ket{\psi}\bra{\psi}$ is the density matrix describing the pure state $\ket{\psi}$ which appears in Eq.~\eqref{eq:rho}.
$g(0)=1$ due to the normalization of $P(\tau)$. 
Then, $g(E)$ should be a continuous function because $\braket{E|\rho|E'}$ and $\braket{E|\chi |E'}$ are continuous functions with respect to $E$ and $E'$.
Furthermore, $g(E)$ should be a positive function because otherwise it contradicts the assumption that there is no zero of $\braket{E|\rho|E'}$.
Using Eq.~\eqref{eq:timing-noise}, the positivity of $g(E)$, the property of the pure state $|\braket{E|\chi|E'}| = \sqrt{\braket{\chi}_E \braket{\chi}_{E'}}$, and $\braket{\chi}_E=\braket{\rho}_E$, we obtain the condition (i) because
\begin{align}
   \frac{|\braket{E|\rho|E'}|}{\sqrt{\braket{\rho}_E \braket{\rho}_{E'}}} 
  &= g(E-E') . \label{eq:g}
\end{align}
The condition (ii) is also satisfied because the Fourier transform of Eq.~(\ref{eq:g}) equals $P(\tau)$ which is a positive function by definition.
The condition (iii) is also satisfied because $e^{i \text{ arg} \braket{E|\rho|E'}}$ can be written in a product form, 
$e^{i \text{ arg} \braket{E|\rho|E'}}
= e^{i \text{ arg} \braket{E|\chi|E'}}
= e^{i \text{ arg}\braket{E|\psi}} e^{i \text{ arg}\braket{\psi|E'}}$.

Now we show that the conditions (i)--(iii) are the sufficient for Eq.~\eqref{eq:rho} with pure state $\chi$.
We choose ansatzes for the pure state $\chi$ and the Fourier transform of the timing noise distribution  as
\begin{align}
\braket{E|\chi_a|E'} 
& \equiv \sqrt{\braket{\rho}_{E} \braket{\rho}_{E'} }
        e^{i \text{ arg} \braket{E|\rho|E'}} , \\
g_a(E-E')
& \equiv \frac{|\braket{E|\rho|E'}|}{\sqrt{\braket{\rho}_E \braket{\rho}_{E'}}} , 
\label{eq:tg}
\end{align}
respectively. Note that the condition (i) is used to define $g_a(E-E')$.
The ansatz satisfies
\begin{equation}
  \braket{E|\rho|E'}  = g_a(E-E') \braket{E| \chi_a|E'}  .  
  \label{eq:timing-noise-ansatz}
\end{equation}
In the same way that Eq.~\eqref{eq:timing-noise} is equivalent to Eq.~\eqref{eq:rho}, Eq.~\eqref{eq:timing-noise-ansatz} is equivalent to 
\begin{equation}
  \rho = \int d \tau \, P_a(\tau) \hat{\mathbf{T}}(\tau) \chi_a \hat{\mathbf{T}}^\dagger (\tau) .
  \label{eq:rho-tP-tchi}
\end{equation}
where $P_a(\tau) \equiv \int d\mathcal{E}\, g_a(\mathcal{E}) e^{-i\mathcal{E}\tau/\hbar}/h$ corresponds to the ansatz for noise distribution $P(\tau)$. 
Eq.~\eqref{eq:rho-tP-tchi} implies that the ansatzes are correct if $P_a(\tau)$ and $\chi_a$ describe a noise distribution and a pure state, respectively.
This is the case because $P_a(\tau)$ is a positive function due to the condition (ii), $\int P_a(\tau) d\tau =1$ due to $g_a(0)=1$ [see Eq.~\eqref{eq:tg}], and $\chi_a$ is a pure state due to the condition (iii).
$\square$

Note that Eqs.~(\ref{eq:timing-noise}) and ~(\ref{eq:g}) imply that the pure state $\chi$ and the timing noise distribution $P(\tau)$ for a given mixed state are unique and determined as,
\begin{align}
  \braket{E|\chi |E'}
  &= \sqrt{\braket{\rho}_{E} \braket{\rho}_{E'} } e^{i \text{ arg} \braket{E|\rho|E'}} , \\
  P(\tau)
  &= \frac{1}{h}\int d E
    \frac{|\braket{\bar{E} +\frac{E}{2} |\rho|\bar{E} -\frac{E}{2}}|}
    {\sqrt{\braket{\rho}_{\bar{E} +\frac{E}{2}} \braket{\rho}_{\bar{E} -\frac{E}{2}}}}
    e^{-i E\tau/\hbar} ,
\end{align}
where $\bar{E} \equiv \int dE\, E \braket{\rho}_E $ is the electron mean energy.

\subsection{In the presence of zero of $\braket{E|\rho|E'}$}
\label{sec:presence-zero}

When the density matrix $\braket{E|\rho|E'}$ has any zero, the derivation in Sec.~\ref{sec:simple} is not valid because $g(E-E')$ in Eq.~\eqref{eq:timing-noise} is not guaranteed to be positive. We show how to generalize the theorem. 

We first observe the property of the possible zeros of state generated by the timing noise. 
Eq.~\eqref{eq:timing-noise}, which is equivalent to Eq.~\eqref{eq:rho}, implies that the zeros of $\braket{E|\rho|E'}$ originate from either $g(E-E') $ or $\braket{E|\chi|E'}$.
The former type of zeros is determined by the character of the timing noise. 
For example when the timing noise distribution $P(\tau)$ is rectangular function with width $w$, its Fourier transform $g(E-E')$ has zeros at $|E-E'| = n h/w $ for integer $n$.
Note that $g(E-E')$ has zeros distributed symmetrically with respect to $E-E'=0$, because $P(\tau)$ and $g(E-E') $ are a real and even function.
We denote $n$th zero of $g(E-E')$ counted from $|E-E'|=0$ as $z_n$, namely 
$g(E-E')=0$ for $|E-E'|=z_n$ and $z_1 <z_2<\cdots$. 
These zeros result in the zeros of the density matrix $\braket{E|\rho|E'}$ spread over diagonals, i.e., $\braket{E|\rho|E'}=0$ along the lines satisfying $|E-E'| = z_n$.

The other type of zeros is determined by the zero of $\braket{E|\chi|E'}$.
These zeros are determined by the character of the pure state $\ket{\psi}$, for example as a result of destructive interference. 
We denote $m$th zero of $\braket{E|\psi}$ as $\zeta_m$, namely $\braket{E|\psi}=0$ for $E = \zeta_m$. 
These zeros result in the zeros of the density matrix $\braket{E|\rho|E'}$ spread over a horizontal and vertical lines which cross at the main diagonal, i.e., $\braket{E|\rho|E'}=0$ for $E = \zeta_m$ or $E'=\zeta_m$.

For the theorem below, we denote the sign $g(\mathcal{E})$ as $\lambda (\mathcal{E})$, namely $\lambda(\mathcal{E})\equiv 1$ when $g(\mathcal{E}) \ge 0$  and $-1$ otherwise.
Due to the property of the zeros of $\rho$, the function $\lambda (\mathcal{E}) $ can be determined by observing the density matrix $\braket{E|\rho|E'}$ and using the following procedure. 
Firstly, $\lambda(0)=1$ because $g(0)=1$ due to the normalization of $P(\tau)$. 
And one can locate the positions of the change of signs of $\lambda(\mathcal{E})$, using the fact that a sign change occurs across the diagonal-type zeros of $\braket{E|\rho|E'}$, namely across $\mathcal{E} = \pm z_n$, which accompany the $\pi$-phase shift in the density matrix $\braket{E|\rho|E'}$.

Now we present the theorem for identifying the timing noise, expressed in terms of the density matrix $\braket{E|\rho|E'}$, the positions of its zeros $z_n$, and the sign function $\lambda$ which can be extracted from the density matrix, hence relying only on the information of the density matrix. The theorem and its derivation are slightly modified from those of Sec.~\ref{sec:simple}.

\textbf{Theorem:}
Eq.~\eqref{eq:rho} with pure state $\chi$ holds if and only if (i) the matrix 
\begin{align}
 \frac{|\braket{E|\rho|E'}|}{\sqrt{\braket{\rho}_E \braket{\rho}_{E'}}} \lambda(E-E')
\end{align}
only depends on energy difference $E-E'$,  (ii) its dependence  has a positive Fourier transform, and
(iii) $e^{i \text{arg} \braket{E|\rho|E'}} \lambda(E-E')$ is rank-1 matrix.

{\it Proof.}
We first show that the conditions (i)--(iii) are necessary for Eq.~\eqref{eq:rho} with pure state $\chi$.
Using Eq.~\eqref{eq:timing-noise}, 
$\lambda(\mathcal{E}) = \text{sgn}[g(\mathcal{E})] $,
the property of a pure state 
$|\braket{E|\chi|E'}| = \sqrt{\braket{\chi}_E \braket{\chi}_{E'}}$, 
and $\braket{\chi}_E = \braket{\rho}_E$,
we find that
\begin{align}
   \frac{|\braket{E|\rho|E'}|}{\sqrt{\braket{\rho}_E \braket{\rho}_{E'}}} \lambda(E-E')
   &= g(E-E') . \label{eq:tg2}
\end{align}
Hence the condition (i) is satisfied.
The condition (ii) is also satisfied because the Fourier transform of Eq.~(\ref{eq:tg2}) equals $P(\tau)$ which is a positive function by definition.
The condition (iii) is also satisfied because
$\lambda(E-E') e^{i \text{ arg} \braket{E|\rho|E'}} = e^{ i \text{ arg} \braket{E|\chi|E'}}$ which is a rank-1 matrix.

Now we show that the conditions (i)--(iii) are sufficient for Eq.~(\ref{eq:rho}) with pure state $\chi$.
We choose ansatzes for the pure state $\chi$ and the Fourier transform of the timing noise distribution  as
\begin{align}
\braket{E|\chi_a|E'} 
&= 
 \sqrt{\braket{\rho}_{E} \braket{\rho}_{E'} } e^{i \text{ arg} \braket{E|\rho|E'}} \lambda(E-E') ,\\
g_a(E-E') 
&= 
\frac{|\braket{E|\rho|E'}|}{\sqrt{\braket{\rho}_E \braket{\rho}_{E'}}} \lambda(E-E') .
\label{eq:tg-gen}
\end{align}
Note that the condition (i) is used to define $g_a(E-E')$.
The ansatzes satisfy
\begin{equation}
  g_a(E-E') \braket{E| \chi_a|E'} = \braket{E|\rho|E'} .  
\end{equation}
Its Fourier transform gives 
\begin{equation}
  \rho = \int d \tau \, P_a(\tau) \hat{\mathbf{T}}(\tau) \chi_a \hat{\mathbf{T}}^\dagger (\tau) .
  \label{eq:rho-tP-tchi-gen}
\end{equation}
Here $P_a(\tau) \equiv \int d\mathcal{E}\, g_a(\mathcal{E}) e^{-i\mathcal{E}\tau/\hbar}/h$ corresponds to the ansatz for noise distribution $P(\tau)$. 
Eq.~\eqref{eq:rho-tP-tchi-gen} implies that the ansatzes are correct if $P_a(\tau)$ and $\chi_a$ describe a noise distribution and a pure state, respectively.
This is the case because $P_a(\tau)$ is a positive function due to the condition (ii), $\int P_a(\tau) d\tau =1$ due to $g_a(0)=1$ [see Eq.~\eqref{eq:tg-gen}], and $\chi_a$ is a pure state due to the condition (iii).
$\square$

Note that Eqs.~(\ref{eq:timing-noise}) and ~(\ref{eq:tg2}) imply that $\chi$ and $P(\tau)$ for a temporally noisy state are unique and determined as,
\begin{align}
  \braket{E|\chi |E'}
  &= \sqrt{\braket{\rho}_{E} \braket{\rho}_{E'} } e^{i \text{ arg} \braket{E|\rho|E'}}
    \lambda(E-E') , \\
  P(\tau)
  &= \frac{1}{h}\int d E
    \frac{|\braket{\bar{E} +\frac{E}{2} |\rho|\bar{E} -\frac{E}{2}}|}
    {\sqrt{\braket{\rho}_{\bar{E} +\frac{E}{2}} \braket{\rho}_{\bar{E} -\frac{E}{2}}}}
    \lambda(E)  e^{-i E\tau/\hbar} ,
\end{align}
where $\bar{E} \equiv \int dE\, E \braket{\rho}_E $ is the mean energy.

\section{Bivariate Gaussian Wigner function}
\label{sec:eNoise}
Here we show an alternative interpretation of bivariate Gaussian Wigner function using the energy noise instead of the timing noise.

We find that a mixed state whose Wigner function is bivariate Gaussian, see Eq.~\eqref{eq:wbg}, can be written as
\begin{equation}
    W(E,t) 
    = \int d\mathcal{E} \tilde{P}(\mathcal{E}) \tilde{W}_0(E-\mathcal{E}, t) .
    \label{eq:W-energyNoise}
\end{equation}
$\tilde{P}(\mathcal{E}) $ is Gaussian distribution describing a random energy shift, instead of the timing shift, with uncertainty $\xi$ given to SES,  
\begin{align}
    \tilde{P}(\mathcal{E}) 
    &= \frac{1}{\sqrt{2\pi} \xi } e^{- \mathcal{E} ^2/(2 \xi^2)} ,
    \\
    \xi
    &= \sqrt{(1-r^2) \Sigma_E^2 - \big(\frac{\hbar}{2 \Sigma_t}\big)^2 } .  
    \label{eq:xi}
\end{align}
We recall that $\Sigma_E$,$\Sigma_t$, and $r$ are energy uncertainty, time uncertainty, and energy-time correlation of $W(E,t)$, respectively.
$\xi$ in Eq.~\eqref{eq:xi} is manifestly real due to the fact that the purity of $W$ is bounded by 1, $\hbar/[2\sqrt{1-r^2}\Sigma_t \Sigma_E] \le 1$.
$\tilde{W}_0$ is a pure-state Wigner function determined as
\begin{equation}
\begin{aligned}
    \tilde{W}_0(E, t) 
    = \frac{2}{h} \exp\Big[& -\big(\frac{1}{2\Sigma_t^2} 
    + \frac{2 r^2 \Sigma_E^2}{\hbar^2} \big) t^2 \\
    &
    + \frac{4 r \Sigma_E \Sigma_t}{\hbar^2}  tE 
    -2 \frac{\Sigma_t^2}{\hbar^2} E^2 \Big] .
\end{aligned}
    \label{eq:tW0}
\end{equation}
One can check easily that the purity of Eq.~\eqref{eq:tW0} is unity, $h\int dE\, dt\, \tilde{W}_0^2(E,t) =1$.
As mentioned in the main text, such alternative interpretation is a consequence of the fact that a bivariate Gaussian Wigner function has energy and time dependence in equal footing;
Eqs.~\eqref{eq:W-energyNoise}--\eqref{eq:tW0} are equivalent to Eqs.~\eqref{eq:W} and \eqref{eq:P-biG}--\eqref{eq:psi-biG} when exchanging the role of energy and time.

\section{Extended results of Fig.~\ref{fig:nCancel_1lev} and \ref{fig:nCancel_TR} }
\label{sec:ext_nCancel}
Here we show how the results of Fig.~\ref{fig:nCancel_1lev} and \ref{fig:nCancel_TR} in the main text extend for different parameters.

\begin{figure}[h]
    \centering    \includegraphics[width=\columnwidth]{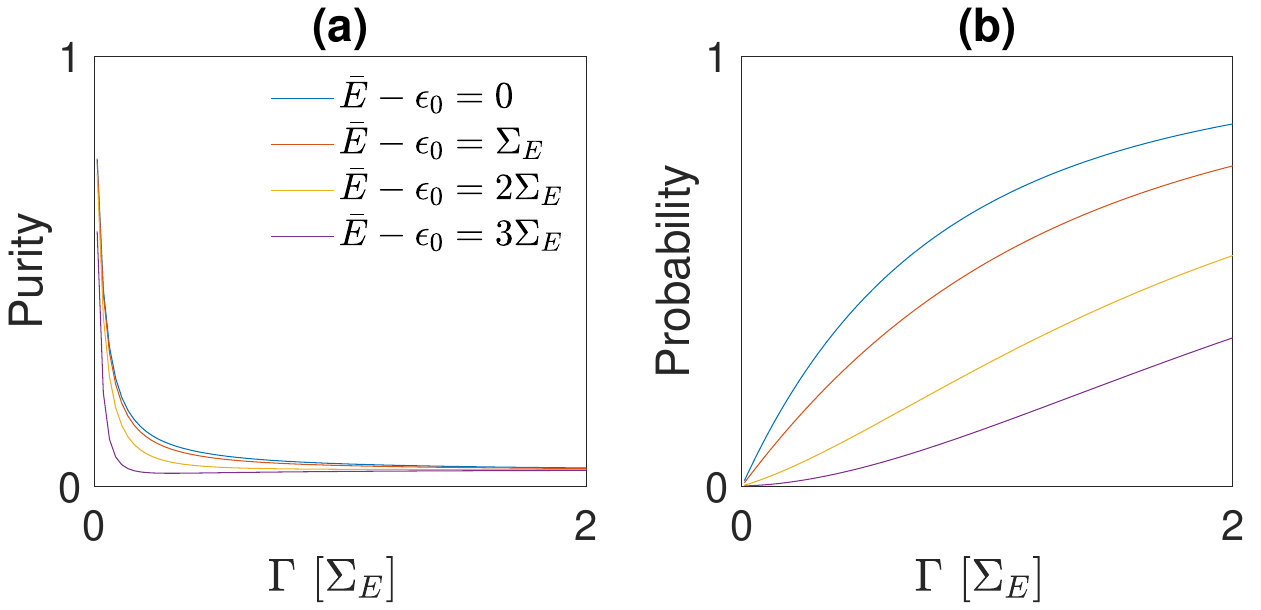}
    \caption{Timing noise cancellation using a resonance level for various mean energy $\bar{E}$ of the packet measured with respect to the resonance level $\epsilon_0$. (a) Purity after the energy filtering through the resonance level. (b) Packet transmission probability. The parameters used for the result are the same as Fig.~\ref{fig:nCancel_1lev} except $\bar{E}-\epsilon_0$. The lines with $\bar{E}-\epsilon_0=0$ (blue) correspond to the result of Fig.~\ref{fig:nCancel_1lev}(b). }
    \label{fig:nCancel_1lev_Ebar}
\end{figure}

Fig.~\ref{fig:nCancel_1lev_Ebar} shows how the result of Fig.~\ref{fig:nCancel_1lev}, where a resonance level is used for the timing noise cancellation, extends when the energy of the resonance level $\epsilon_0$ is not aligned with the packet mean energy $\bar{E}$. In this case, the transmission probability through the resonance level is $\mathbb{T}(E) = 1/[1+(E-\epsilon_0)^2/\Gamma^2]$. The result shows that as the resonance level deviates from the packet mean energy, both the filtered purity and the probability both decrease. This indicates that it is best to align the resonance level with the packet mean energy for the timing noise cancellation.
However, as long as the resonance level is near the mean energy, $\epsilon_0 \in [\bar{E} -\Sigma_E, \bar{E}+\Sigma_E]$, the timing noise cancellation is achieved with almost the same results. 

\begin{figure}[h]
    \centering
    \includegraphics[width=\columnwidth]{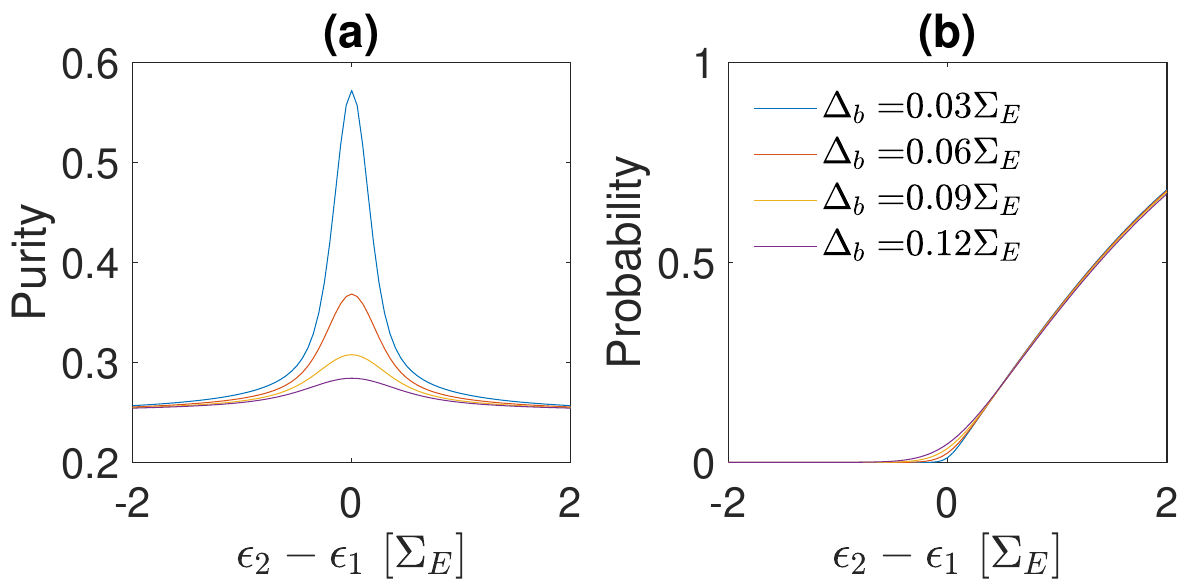}
    \caption{Timing noise cancellation using the potentical barriers for various broadening $\Delta_b$. (a) Purity after the energy filtering through the barriers. (b) Post-selection probability. The parameters used for the results are the same as Fig.~\ref{fig:nCancel_TR} except $\Delta_b$. The lines with $\Delta E_b =0.03 \Sigma_E$ (blue) correspond to the result of Fig.~\ref{fig:nCancel_TR}(b).  }
    \label{fig:nCancel_barriers_dEbs}
\end{figure}

Fig.~\ref{fig:nCancel_barriers_dEbs} shows how the result of Fig.~\ref{fig:nCancel_TR}, where a potential barriers are used for the timing noise cancellation, extends when the broadening of the barrier, $\Delta_b$, is varied. As $\Delta_b$ increases, the purity decreases while the probability slightly increases. This indicates that it is best to use the barriers with small $\Delta_b$, which corresponding to thick barrier, for the timing noise cancellation. And as long as $\Delta_b$ is much smaller than the energy uncertainty $\Sigma_E$, the timing noise cancellation is achieved with similar results.

\end{document}